\documentclass[
    ,final            % use final for the camera ready runs
  ,numberedheadings % uncomment this option for numbered sections
  ]
  {aipproc}

\usepackage{amsmath}
\usepackage{amssymb}

\layoutstyle{8x11double}

\begin{document}

\newcommand{\blue}[1]{\textcolor{blue}{#1}}

\newcommand{\beq}{\begin{equation}}
\newcommand{\eeq}{\end{equation}}
\newcommand{\beqa}{\begin{eqnarray}}
\newcommand{\eeqa}{\end{eqnarray}}
\newcommand{\bmat}{\begin{displaymath}}
\newcommand{\emat}{\end{displaymath}}

\newcommand{\eq}[1]{Eq.~(\ref{#1})}

\newcommand{\lan}{\langle}
\newcommand{\ran}{\rangle}

\title{Rigidity of glasses and jamming systems at low temperatures}

%                \texttt{http://www.aip..org/pacs/index.html}>

%Flasticity, mechanical properties of solids, 62.20.D-
%Glasses 61.43.Fs
%Disordered Solids 61.43.-j
%Glasses, rheology, 83.80.Ab
% Glass transitions, theory and modeling of, 
%Colloids glass transition 64.70.pv
%Colloids rheology of 83.80.Hj
%Emulsions, rheological properties, 83.80.Iz

\author{Hajime Yoshino}{
  address={Department of Earth and Space Science, School of Science,
 Osaka University, Toyonaka 560-0043, Japan.}
}

%\author{<author2>}{
%  address={<common address for author2 and author3>}
%}

%\author{<author3>}{
%  address={<common address for author2 and author3>}
%  ,altaddress={<author1 address>} % additional visiting address
%}

\begin{abstract}
We discuss a microscopic scheme to compute the rigidity of glasses
 or the plateau modulus of supercooled liquids by twisting replicated liquids.
We first summarize the method in the case of harmonic glasses with
analytic potentials. Then we discuss how it can be extended 
to the case of repulsive contact systems : the hard sphere glass and related 
systems with repulsive contact potentials which enable the jamming
 transition at zero temperature. 
For the repulsive contact systems we find  entropic rigidity 
which behaves similarly as the pressure in the low temperature limit:
it is proportional to the temperature
$T$ and tends to diverge approaching the jamming density $\phi_{\rm J}$
with increasing volume fraction $\phi$ as $\lim_{T \to 0} \mu/T \propto
 1/(\phi_{\rm J}-\phi)$, which may account for experimental observations 
of rigidities of repulsive colloids and emulsions.
\end{abstract}

\maketitle

%%%%%%%%%%%%%%%%%%%%%%%%%%%%%%%%%%%%%%%%%%%%
%% MAINMATTER
%%%%%%%%%%%%%%%%%%%%%%%%%%%%%%%%%%%%%%%%%%%%

\section{Introduction}

\begin{figure}[htbp]
%\begin{center}
\includegraphics[width=0.3\textwidth]{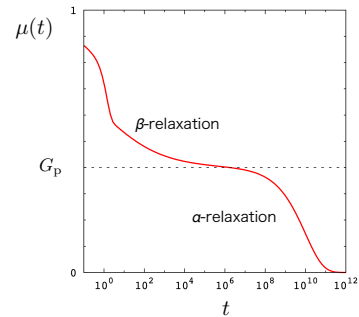}
%\end{center}
\caption{Schematic picture of stress relaxation in supercooled liquids.
A simplest protocol is considered: switch on the
shear-strain of amplitude $\gamma$ (See Fig.~\ref{fig-shear-geometry})
at time $t=0$ and measure the relaxation of shear-stress $\sigma(t)$ 
which follows. Here we define $\mu(t)=\sigma(t)/\gamma$. 
The plateau value $G_{\rm p}$ of $\mu(t)$ is called as the plateau modulus
which represents the effective rigidity or 
shear-modulus of metastable glassy states.
}
\label{fig-two-step-func-model}
\end{figure}

Supercooled liquids, glasses and jamming systems exhibit rich
visco-elasticity \cite{Angell-Ngai-McKenna-McMillan-Martin}: 
at shorter time scales called as the
$\beta$-regime such a system behaves as a solid with finite rigidity 
while it behaves as a liquid with high viscosity at longer time scales called
as the $\alpha$-regime. These features appear clearly 
in the relaxation of shear-stress (see Fig.~\ref{fig-two-step-func-model}) 
which follows after switching on a small shear-strain $\gamma$ (See
Fig.~\ref{fig-shear-geometry}). Approaching the glass transition point
the separation of the time scales between the two regimes become enormous
so that a supercooled liquid behaves essentially as a quasi-static solid for a
very long time. The shear-modulus or the rigidity is the most basic 
quasi-static property  which distinguishes solids from other states of matters.

Among the various types of glasses a class of 
systems like densely packed repulsive colloids, emulsions, foams 
and granular particles \cite{Weeks}
exhibit an interesting common feature called as the jamming transition: 
the characters of the amorphous solid state change around the so called
jamming point at a certain volume fraction $\phi_{\rm J}$ at zero temperature. 
This is manifested in various quasi-static as well as certain dynamic 
properties of such amorphous solids 
\cite{Durian-1995,Mason-Weitz-1995,Mason-Bibette-Weitz-1995,Weitz97,Nagel-group,Wyart-2005,Dauchot-2005,Brito-Wyart-2006,berthier-witten-2009,hecke-2010,parisi-zamponi-2010,Berthier-Jacquin-Zamponi-2011,Schreck-et-al-2011,Wyart-2012,ikeda-berthier-biroli-2012,Otsuki-Hayakawa-2012,Coulais-2012,Olsson-Teitel,Hatano,Otsuki-Hayakawa,Ikeda-Berthier-Sollich-2012}.

The wide time separation between the two regimes suggest that the
quasi-static responses in the $\beta$-regime may be analyzed by some
statistical mechanical approaches such as the cloned liquid approach
which combines the replica method and liquid theory 
\cite{mezard-parisi-1999,coluzzi-mezard-parisi-verrocchio-1999,
parisi-zamponi-2010,berthier-private-communication-2012}.
The latter is a first principle, microscopic approach within
the framework of the so called random first order transition (RFOT)
theory \cite{RFOT,RFOT-review}.
Indeed we showed recently that the replica method provides a trick
to {\it disentangle} the $\beta$ and $\alpha$-like responses and 
extract the quasi-static part of the responses 
in supercooled liquids and structural glasses \cite{paper1,paper2}.

The purpose of the  the present paper is two fold: we first review the method
\cite{paper1,paper2} developed for systems with analytic potentials
such as the Lennard-Jones potential which are suited for  usual 
molecular glasses.
These systems behave as harmonic solids at low temperatures,  i.~e.
systems of random spring networks.
Then we attempt to extend the method to account for systems which exhibit
the jamming transition. The essential ingredient in such systems
is the repulsive contact potentials such as hard-spheres and 
some soft repulsive contact potentials which are not analytic.
Based on this method we analyze the behaviour of the rigidity of hard-spheres and 
soft repulsive contact systems in the low temperature limit
approaching the jamming density $\phi_{\rm J}$ from below. It appears that our result
accounts for some experimental observations of the rigidity of repulsive colloids and emulsions \cite{Mason-Weitz-1995,Mason-Bibette-Weitz-1995,Weitz97}.

The organization of the paper is as follows. 
In the next section we introduce the two distinct classes of systems:
systems with the analytic potentials and the repulsive contact potentials.
In the subsequent sections we review our strategy  \cite{paper1,paper2}
to extract the quasi-static response functions 
of supercooled liquids and glasses
based on the cloned liquid approach.
Then we review the basic fluctuation formulae of the rigidity and  
our previous scheme to compute the rigidity of harmonic glassy systems
reported in \cite{paper1,paper2}. Finally we discuss the extension 
of the method to the cases of repulsive contact potentials and analyze how the
jamming transition is reflected on the {\it entropic} 
rigidity approaching $\phi_{\rm J}$.

\section{Models}

We consider a generic system of $N$ particles
($i=1,2,\ldots,N$) in the 3-dimensional space with volume $V$
interacting with each other through
a two-body potential $v(r)$ which only depends on the relative distance
$r$ between particles.
The potential part of the Hamiltonian can be written as,
\beq
U = \sum_{\langle i j \rangle} v(r_{ij}) \qquad 
r_{ij}=|{\bf r}_{i}-{\bf r}_{j}|
\eeq
where $\langle i j \rangle$ stands for summation over the $N(N-1)/2$ pairs
of the particles and ${\bf r}_{i}$ ($i=1,2,\ldots,N$) 
represents the position of the
particles. We suppose that the temperature $T$ is low enough
and the number density $\rho=N/V$ is high enough such that the system
is in a supercooled liquid or a glassy metastable state.

We consider two distinct classes of systems:
\begin{itemize}
\item {\it Harmonic systems}: the potential $v(r)$ 
is an analytic function of $r$ for $r>0$
like the Lennard-Jones potential. Presumably this class of systems is
relevant for molecular glasses. 

For an explicit model computation
we consider the soft-core potential $v(r)=\epsilon(r/a)^{-12}$
where $\epsilon$ and $a$ are the unit energy and length respectively.

\item {\it Repulsive contact systems} : the potential $v(r)$ is repulsive and
has a definite cut-off at the scale the particle size $a$ 
like the hard-spheres.
Presumably this class of systems is relevant for emulsions 
and repulsive colloids.
The density is a crucial parameter in these systems and 
it is convenient to represent it 
via volume fraction $\phi$ which is related to the number density $\rho$
as $\phi=(\pi/6)\sigma^{3}\rho$.
An important feature of this class of systems is that they exhibit
the jamming transition at $T=0$ by increasing the volume fraction
$\phi$ up to  some jamming density $\phi_{\rm J}$.

For an explicit model computation we consider the soft-particle
potential $v(r)=\epsilon(1-r/a)^{2}\theta(1-r/a)$ where
$\theta(r)$ is the step function.  In the present paper we
limit ourselves to the range of volume fractions $\phi < \phi_{\rm J}$
where the system behaves as hard-spheres in the zero
temperature limit $T \to 0$.

\end{itemize}

\section{Disentanglement of the intra-state and inter-state responses}

\subsection{A mean-field picture: ensemble of metastable states}

Let us take the basic energy landscape picture
\cite{Stillinger-Weber-1982,Doliwa-Heuer-2003} :
we consider that the equilibrium state of a supercooled liquid or a glass can be described
in terms of a statistical ensemble of metastable states,
which might be interpreted as metabasins \cite{Doliwa-Heuer-2003} each
of which is a union of inherent structures \cite{Stillinger-Weber-1982}.
 Let us label the metastable states
as $\alpha=1,2,3,\ldots$ and denote the free-energy per particle of the $\alpha$-th state as $f_{\alpha}$.
Then the equilibrium free-energy of the system may be expressed formally as,
\begin{eqnarray}
F(h)=-k_{\rm B}T\log Z(h) \nonumber \\
Z(h) \simeq \sum_{\alpha}e^{-N \beta f_{\alpha}(h)}
\end{eqnarray}
where $k_{\rm B}$ is the Boltzmann's constant and $\beta \equiv 1/k_{\rm B}T$.

We have introduced a parameter $h$ which represents a generic infinitesimal probing field, such as the shear 
which we will focus on in the present paper. 
The linear susceptibility to the external field $h$ can be seen to
take the following generic form,
\beq
\chi\equiv \left. \frac{1}{N}\frac{d^{2} F(h)}{dh^{2}} \right |_{h=0}
=\hat{\chi}+\tilde{\chi}.
\label{eq-chi-decomposition}
\eeq
with
\beq
\hat{\chi}=[\![ \chi_{\alpha} ]\!]
\qquad \tilde{\chi}=\beta N ( [\![ o^{2}_{\alpha}]\!]
-[\![ o_{\alpha}]\!]^{2})
\label{eq-hat-chi-tilde-chi}
\eeq
Here $o_{\alpha} \equiv \left. df_{\alpha}(h)/dh \right |_{h=0}$
is the equilibrium value of an observable $o$ which is conjugated to the 
external field $h$. Similarly $\chi_{\alpha} \equiv d o_{\alpha}/dh=\left. d^{2}f_{\alpha}(h)/dh^{2} \right |_{h=0}$
is the associated linear susceptibility {\it within a given metastable
state} $\alpha$. In \eqref{eq-hat-chi-tilde-chi}
$[\![ \ldots ]\!]$ stands for averaging over the ensemble of the metastable states defined as,
\beq
[\![ \ldots ]\!] \equiv   \frac{\sum_{\alpha} e^{-N \beta f_{\alpha}(0)} \ldots}{Z(0)}.
\eeq

The important feature evident in \eqref{eq-chi-decomposition}
is that the total linear susceptibility $\chi$
is the sum of two distinct parts: $\hat{\chi}$ and $\tilde{\chi}$ associated with the response within metastable states and response due to jumps between different metastable states. Physically $\hat{\chi}$ can be
regarded as the quasi-static response within the $\beta$-regime and $\tilde{\chi}$ can be 
related to the response in the $\alpha$-regime.
Although the two parts have very different characters, 
\eqref{eq-chi-decomposition} implies they are mixed up in the total response.
We wish to disentangle the two. Let us discuss below how the replica trick
works  for this purpose.

\subsection{Response of a cloned system}

Let us consider a {\it cloned system}
\cite{monasson-1995,Franz-Parisi,mezard-parisi-1999} 
which consists of $m$ replicas of
the same system labeled as $a=1,2,\ldots,m$. The free-energy of the cloned system is defined as, 
\begin{eqnarray}
F_{m}(\{h_{a}\})=-\frac{k_{\rm B}T}{m} \log Z_{m}(\{h_{a}\}) \nonumber \\
Z_{m}(\{h_{a}\}) \simeq \sum_{\alpha}e^{-N \beta
 \sum_{a=1}^{m}f_{\alpha}(h_{a})} 
\label{eq-free-ene-cloned}
\end{eqnarray}
Note that there is only one summation over the metastable states
instead of $m$ summations. This means that we are assuming that 
$m$ replicas  are not allowed to fluctuate independently from each other
but forced to fluctuate together over different metastable states.
Yet the replicas are allowed to fluctuate differently from each other 
within the metastable states. How to realize such a situation in practice 
is a non-trivial task by itself \cite{monasson-1995,Franz-Parisi,mezard-parisi-1999,parisi-zamponi-2010} as we discuss shortly later. 

The key point is that we have put different probing fields $h_{a}$
($a=1,2,\ldots,m$)  on different replicas in \eqref{eq-free-ene-cloned} \cite{paper1,paper2}.
It naturally lead us to define a sort of generalized 
linear-susceptibility of a matrix form,
\beq
\chi_{ab}\equiv \left. \frac{1}{N}\frac{\partial^{2} F_{m}(\{h\})}{\partial h_{a}\partial h_{b}}\right |_{\{h_{a}=0\}}
=\hat{\chi}_{m}\delta_{ab}+ \tilde{\chi}_{m}.
\label{eq-chi-ab}
\eeq
where $\hat{\chi}_{m}$ and $\tilde{\chi}_{m}$ are almost the same as
$\hat{\chi}$ and $\tilde{\chi}$ defined in \eqref{eq-hat-chi-tilde-chi}
but evaluated by replacing $[\![\ldots ]\!]$ by $[\![\ldots ]\!]_{m}$
defined as,
\beq
[\![ \ldots ]\!]_{m} \equiv 
\frac{\sum_{\alpha}  e^{-N m \beta f_{\alpha}(0)}}{Z_{m}(0)}.
\eeq
Quite remarkably the 2nd equation
of \eqref{eq-chi-ab},  which can be easily verified, 
implies that the $\beta$ and $\alpha$-like responses can be distinguished 
from each other: 
\begin{eqnarray}
\hat{\chi}=\lim_{m \to 1}\hat{\chi}_{m} \qquad
 \hat{\chi}_{m}=\chi_{aa}-\chi_{a\neq b}  \\
 \tilde{\chi}= \lim_{m \to 1}\tilde{\chi}_{m}  
\qquad \tilde{\chi}_{m}=\chi_{a\neq b}
\label{eq-chi-m-1}
\end{eqnarray}

\subsection{Cloned liquid}

Here let us briefly sketch how to implement a cloned system \cite{mezard-parisi-1999,parisi-zamponi-2010}.
The basic idea is to introduce a system of an artificial {\it molecular liquid} 
in which each 'molecule' $i=1,2,\ldots,N$ consists of $m$
particles belonging to different replicas $a=1,2,\ldots,m$.
The particles are allowed to fluctuate only within the molecule of
size $A$, which is interpreted physically as the {\it cage size}.
The cage size $A$ is determined by a variational principle (see below).
Existence of a solution with a finite cage size $A < \infty$ 
implies existence of metastable states  \cite{Franz-Parisi,mezard-parisi-1999}.

The coordinates of the particles ${\bf r}^{a}_{i}$ ($i=1,2,\ldots,N$)
can be decomposed formally as,
\beq
{\bf r}^{a}_{i}={\bf R}_{i}+{\bf u}_{i}^{a} \qquad
{\bf R}_{i}\equiv \frac{1}{m}\sum_{a=1}^{m}{\bf r}^{a}_{i}
\label{eq-decomposition}
\eeq
where ${\bf u}_{i}^{a}$ stands for fluctuation 
of the particle belonging to  the $a$-th replica
with respect to the center of mass ${\bf R}_{i}$ of the molecule. 
The fluctuations within the molecules
are assumed to obey the Gaussian statistics 
with the mean and variance given by \cite{mezard-parisi-1999}, 
\begin{equation}
\langle ({\bf u}_{i}^{a})^{\mu} \rangle_{\rm cage}=0 \qquad 
\langle ({\bf u}^{a}_{i})^{\mu}({\bf u}^{b}_{j})^{\nu} \rangle_{\rm cage} = A(\delta_{ab}-\frac{1}{m})\delta_{\mu \nu}\delta_{ij}
\end{equation}
Here $\mu$ (and $\nu$)
represents a component of 3-dimensional vectors $\mu=x,y,z$.
The factor $\delta_{ab}-\frac{1}{m}$ reflects the constraint
$\sum_{a=1}^{m}({\bf u}^{a})={\bf 0}Z$.

The free-energy of the molecular or cloned liquid $G_{m}(A)$ of
a given number of replicas $m$ and the cage size $A$ can be 
obtained as follows \cite{mezard-parisi-1999}. First one integrate out 
the fluctuations within the molecules which amounts to replace
the original interaction potential $v(r)$ by a remornalized one $v_{\rm
eff}(r, A)$. \cite{mezard-parisi-1999,parisi-zamponi-2010} 
In the case of analytic potentials it reads as \cite{mezard-parisi-1999},
\beq
v_{\rm eff}(r)=v(r)-(1-m)\frac{A}{m^{2}}\nabla^{2}v(r)+\ldots.
\label{eq-veff}
\eeq
Then one is left to integrate out the CM positions of the molecules interacting with 
each other via $v_{\rm eff}(r, A)$ and subjected to a heat-bath
at an effective temperature $T/m$. 
Eventually we  have to take the $m \to 1$ limit (see \eqref{eq-chi-m-1}).

The strategy is to start from sufficiently small $m (\ll 1)$
so that the cloned system remain in the liquid state 
because the effective temperature $T/m$ becomes sufficiently high
even if the actual temperature $T$ itself is very low.
Then standard density functional methods
of the liquid theory \cite{hansen-mcdonald} 
allows one to compute the free-energy $G_{m}(A)$. 
The value of $A$ is determined by 
minimizing the variational free-energy $G_{m}(A)$ with respect to $A$ 
yielding 
$F_{m}={\rm min}_{A}G_{m}(A)$. Let us denote the value of the cage size $A$ at the minimum
as $A^{*}(m,T,\rho)$.

The last step is to take the limit $m \to 1^{-}$. 
It turns out that at temperatures $T$ below 
the ideal glass transition temperature, i. e.
the Kauzmann temperature $T_{\rm K}(\rho)$, one finds
a characteristic value $m^{*}=m^{*}(T,\rho)$ in the range
$0 \leq  m^{*}(T,\rho) \leq 1$
such that 
\beq
\lim_{m \to 1^{-}} F_{m}(T,\rho)=F_{m^{*}}(T,\rho).
\label{eq-f-m-1}
\eeq
holds. The Kauzmann transition temperature $T_{\rm K}(\rho)$ can be obtained 
by solving $m^{*}(T_{\rm K}(\rho),\rho)=1$. 
The reason behind \eqref{eq-f-m-1} is actually 
the entropy crisis mechanism, i.~e. the ideal glass transition, taking place along
the $m$-axis. We refer the readers to Refs \cite{mezard-parisi-1999} for the details.
The above observation implies 
that $\beta$ and $\alpha$-like responses 
(see \eqref{eq-chi-m-1}) can be obtained as,
\beq
\hat{\chi}=\hat{\chi}_{m^{*}}
 \qquad \tilde{\chi}=\tilde{\chi}_{m^{*}}.
\eeq

\section{Microscopic computation of the rigidity of glasses}

\subsection{Static linear response to shear}

\begin{figure}[htbp]
%\begin{center}
\includegraphics[width=0.3\textwidth]{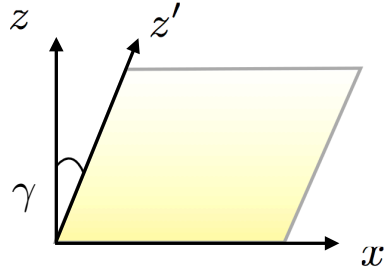}
%\end{center}
\caption{The geometry of a sheared container.}
\label{fig-shear-geometry}
\end{figure}

Now let us study static linear response to shear. 
For clarity we recall and compare the well known fluctuation formulae for the shear-modulus
for analytic potentials \cite{squire} and hard-spheres \cite{Farago-Kantor-2000}
in the case of simple shear. 
To this end we consider a system of particles $i=1,2,\ldots,N$ 
put in a rectangular 
container of volume $V$ and perturb the system by 
a shear-strain of {\it infinitesimal} amplitude $\gamma$ 
on the container (See Fig.~\ref{fig-shear-geometry}). 
Most important feature of the 
{\it shear} is that it just changes the {\it shape} of
the container but not the volume $V$ (and thus the density $\rho$).

The free-energy $F(\gamma)$ of
the system may be formally expanded in power series of $\gamma$ as,
\beq
F(\gamma)/V=F(0)/V+\gamma \sigma + \frac{\gamma^{2}}{2}\mu + \ldots.
\eeq
Here the coefficients of the 1st and 2nd order terms in the expansion
defines the shear-stress $\sigma$ and the rigidity or the shear-modulus
$\mu$. Microscopic expressions of $\sigma$ and $\mu$ can be obtained as follows.

The free-energy $F(\gamma)$ can be expressed formally as,
\begin{eqnarray}
&& F(\gamma) \equiv -k_{\rm B} T \log Z(\gamma) \nonumber \\
&& Z(\gamma)\equiv \int_{{\cal V}(\gamma)} \prod_{i=1}^{N}  \frac{d^{3}r_{i}}{\lambda_{\rm th}^{3}}
e^{-\beta \sum_{\langle i j \rangle}v(r_{ij})}\label{eq-F-gamma} \\
&& =\int_{{\cal V}(0)} \prod_{i=1}^{N}  \frac{d^{3}r'_{i}}{\lambda_{\rm th}^{3}}
\left. e^{-\beta \sum_{\langle i j \rangle}  v(r_{ij})} \right |_{r_{ij}=\sqrt{(x'_{ij}+\gamma
z'_{ij})^{2}+(y'_{ij})^{2}+(z'_{ij})^{2}}}
\nonumber
\end{eqnarray}
Here $\lambda_{\rm th}$ is the thermal de Brogile length.
The subscripts ${\cal V}(\gamma)$ represent the range of integrations, 
including not only the volume $V$ (which is invariant under shear) 
but also its {\it shape}, parametrized by the shear-strain $\gamma$.
In the 2nd equation we changed the integration variables 
from ${\bf r}$ to ${\bf r}'$ (See Fig.~\ref{fig-shear-geometry})
which allows us to change the integration region ${\cal V}(\gamma)$
back to the unperturbed one ${\cal V}(0)$. The 2nd equation allows us
to easily obtain the expansion of the free-energy $F(\gamma)$ in power
series of $\gamma$. 

\paragraph{Shear-stress}

Microscopic expression of the shear-stress is obtained as,
\beq
\sigma \equiv\left. \frac{1}{V}\frac{\partial F(\gamma)}{\partial \gamma} \right |_{\gamma=0}=\frac{1}{V}\sum_{\langle ij \rangle} \langle \sigma({\bf r}_{ij}) \rangle
\label{eq-def-sigma}
\eeq
where $\langle \ldots \rangle$ is the thermal average,
\beq
\langle \ldots \rangle \equiv \frac{1}{Z(0)}
\int_{{\cal V}(0)} \prod_{i=1}^{N}  \frac{d^{3}r'_{i}}{\lambda_{\rm th}^{3}}
 e^{-\beta \sum_{\langle i j \rangle}  v(r_{ij})} \ldots.
\eeq
Here we introduced the {\it local shear-stress},
\beq
\sigma({\bf r}) \equiv \hat{z}\hat{x}r v'(r)
\label{eq-sigma-harmonic}
\eeq
with short-hand notations 
$\hat{x} \equiv x/r$, $\hat{y} \equiv y/r$ ,  $\hat{z} \equiv z/r$
where $r=\sqrt{x^{2}+y^{2}+z^{2}}$. 

Remark: note that the expression \eqref{eq-sigma-harmonic}
is ill-defined for hard-spheres.
However the average shear-stress \eq{eq-def-sigma} can still be
evaluated safely by noting that 
$ -\beta v'(r)e^{-\beta v(r)}=(e^{-\beta v(r)})'$ becomes 
a delta function $\delta(r-a)$  for hard-spheres
and soft repulsive contact systems in the $T \to 0$ limit. 

\paragraph{Shear-modulus}

Similarly the microscopic expression or
the static fluctuation formula of the shear-modulus is obtained as,
\begin{eqnarray}
&& \mu\equiv\left. \frac{1}{V}\frac{\partial^{2} F(\gamma}{\partial \gamma^{2}} \right |_{\gamma=0}  =\mu_{\rm born} \label{eq-mu-harmonic}\\
&& -\beta V \sum_{\langle kl \rangle}\sum_{\langle mn \rangle} 
(\langle \sigma({\bf r}_{kl}) \sigma({\bf r}_{mn})\rangle 
-\langle \sigma({\bf r}_{kl})\rangle\langle \sigma ({\bf r}_{mn}) \rangle) 
\nonumber 
\end{eqnarray}
where $\mu_{\rm born}$ is the so called Born term defined as,
\beq
\mu_{\rm born}\equiv
\frac{1}{V}\sum_{ \langle i j \rangle}
\hat{z}^{2}  [
 r^{2}v''(r)
\hat{x}^{2}
+  r v'(r)
(1- \hat{x}^{2}) ].
\label{eq-mu-born}
\eeq
The Born term $\mu_{\rm born}$ represents the instantaneous, affine response to shear
while the 2nd term in the r.h.s of the 2nd equation \eqref{eq-mu-harmonic}
represents the so called non-affine correction 
due to stress relaxation \cite{squire,paper2}. 

The above expressions \eqref{eq-mu-harmonic} \eqref{eq-mu-born} are problematic
for the repulsive contact systems. 
Especially the Born term is formally infinite for hard-spheres.
The Born term \eq{eq-mu-born} stems from 
direct spatial derivatives of the local stress \eqref{eq-sigma-harmonic} 
which does not exist in this class of systems in sharp 
contrast to the harmonic systems (See the remark below 
\eqref{eq-sigma-harmonic}).
Then for this class of systems it is more convenient 
to use an alternative but equivalent expression \cite{Farago-Kantor-2000},
\begin{eqnarray}
&& \beta \mu= V \left [\sum_{\langle kl\rangle} \langle \beta \sigma({\bf r}_{kl})\rangle^{2} \right . \label{eq-mu-hardsphere} \\
&& \left. -\sum_{\langle kl \rangle}\sum_{\langle mn \rangle \neq \langle kl \rangle} 
(\langle \beta \sigma({\bf r}_{kl}) \beta \sigma({\bf r}_{mn})\rangle 
-\langle \beta \sigma({\bf r}_{kl})\rangle\langle \beta \sigma({\bf r}_{mn})\rangle)
\right. ]
\nonumber
\end{eqnarray}
which can also be obtained from \eqref{eq-F-gamma}.
In the derivation one has to
perform some integrations by parts in order to get rid of the Born term.
In \eqref{eq-mu-hardsphere} we dropped off some terms
which cancel out with each other exactly in isotropic systems. 
This assumption is valid for the systems we consider below.

In liquids the terms on the r.h.s of \eqref{eq-mu-harmonic}\eqref{eq-mu-hardsphere}
cancel with each other to realize $\mu=0$ which reflects
the translational invariance of liquids. 

\section{Harmonic systems}

Let us now discuss the rigidity of the glassy states of the harmonic systems whose potentials $v(r)$ are analytic for $r>0$. Here we present a summary of the results reported \cite{paper1,paper2} which will be compared with the case of the repulsive contact systems analyzed in the next section. 
First we consider a 'free' $m$-replica system without any 'cloning': the $m$ replicas
are totally independent from each other.
For such a system we can naturally define a rigidity matrix $\mu_{ab}$
and find its microscopic expression similarly as
to the one for the single system \eqref{eq-mu-harmonic},
\begin{eqnarray}
&&\mu_{ab}\equiv \frac{1}{V}\frac{\partial^{2} F_{m}}{\partial \gamma_{a}\gamma_{b}} 
=\mu_{\rm born}\delta_{ab}\nonumber\\
&&-\beta V \sum_{\langle kl \rangle}\sum_{\langle mn \rangle} 
(\langle \sigma({\bf r}^{a}_{kl}) \sigma({\bf r}^{b}_{mn})\rangle 
-\langle \sigma({\bf r}^{a}_{kl})\rangle\langle \sigma ({\bf r}^{b}_{mn}) \rangle),
\nonumber 
\end{eqnarray}
where $\mu_{\rm born}$  and the local shear-stress $\sigma({\bf r})$
are the same as those defined in \eq{eq-mu-born} and \eq{eq-sigma-harmonic}.

Now we switch on the {\it cloning} following the prescription discussed previously
using the decomposition of the coordinates \eqref{eq-decomposition}.
This allows us to expand the local shear-stress $\sigma({\bf r}_{ij}^{a})$ as,
\beq
\sigma({\bf r}_{ij}^{a})=\sigma({\bf R}_{ij})
+\left. \nabla \sigma(r) \right |_{r={\bf R}_{ij}} 
\cdot ({\bf u}_{i}^{a} -{\bf u}_{j}^{a})+ \ldots
\nonumber 
\eeq
as well as the interaction potential as,
\begin{eqnarray}
&& v(r_{ij}^{a})=v(R_{ij})+\nabla \left. v(r) \right |_{r=R_{ij}} \cdot 
({\bf u}_{i}^{a} -{\bf u}_{j}^{a}) \nonumber \\
&&+\frac{1}{2}\left. \sum_{\mu\nu}\frac{\partial^{2}v(r)}{\partial r^{\mu}\partial r^{\nu}}
\right |_{r=R_{ij}} ((u_{i}^{a})^{\mu} -(u_{j}^{a})^{\mu})
((u_{i}^{a})^{\nu} -(u_{j}^{a})^{\nu}) + \ldots. 
\nonumber
\end{eqnarray}

As the result of the cloning we obtain the rigidity matrix in the form
of the anticipated matrix structure \eqref{eq-chi-ab},
\beq
\mu_{ab}=\hat{\mu}\delta_{ab}+\tilde{\mu} 
\eeq
with the rigidity of metastable states (plateau modulus $G_{\rm p}$) obtained up to $O(A)$ as,
\begin{eqnarray}
&& \hat{\mu}=  \mu^{\rm eff}_{\rm born}  \label{eq-hat-mu-harmonic}\\
&& -\frac{k_{\rm B}T}{m^{*}}\left (\frac{A^{*}}{m^{*}} \right)\rho 
%\left[ 
 \int d^{3}r |\left (\beta m^{*} \nabla \sigma({\bf r}) \right)|^{2}
g_{T/m^{*},\rho}({\bf r}) 
%\right. 
\nonumber
\\
&& 
-\frac{k_{\rm B}T}{m^{*}}\left (\frac{A^{*}}{m^{*}} \right)\rho 
%\left. 
\int d^{3}r_{1}d^{3}r_{2} \nonumber \\
&& \left (\beta m^{*} \nabla \sigma({\bf r}_{1}) \right)
\cdot  \left (\beta m^{*} \nabla \sigma({\bf r}_{2}) \right)
(g_{3})_{T/m^{*}}({\bf r}_{1},{\bf r}_{2})
% \right ]
\nonumber
\end{eqnarray}
while the $\alpha$-like part of the response is given by
\beq
\tilde{\mu}=-\frac{\hat{\mu}}{m^{*}}
\eeq
because of the sum rule $\sum_{b=1}^{m^{*}}\mu_{ab}=0$ reflecting the plain fact
the cloned liquid as a whole is just a liquid. Physically the latter
suggests static analogue of the yielding processes (See  \cite{paper2} for discussions).
In the above equations $m^{*}=m^{*}(T,\rho)$ and $A^{*}=A^{*}(m^{*}(T,\rho),T,\rho)$ are the values
determined in the course of the evaluation of the free-energy of the
cloned liquid discussed before.
Here $g_{T,\rho}(r)$ and $(g_{3})_{T,\rho}({\bf r}_{1},{\bf r}_{2})$ 
are the radial distribution
function and the three-point correlation function of the liquid at
temperature $T$ and density $\rho$ respectively.

\begin{figure}[htbp]
%\begin{center}
\includegraphics[width=0.3\textwidth]{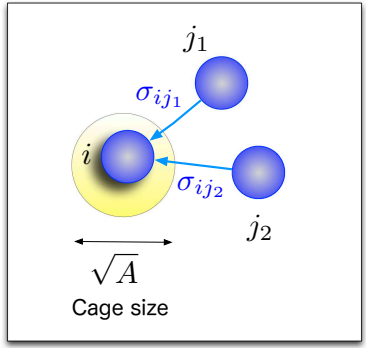}
%\end{center}
\caption{Schematic picture of stress relaxation due to motions of
particles inside cages. The local motion of the particle $i$ can relax
local stress between the particle $i$ and the surrounding particles $j_{1},j_{2},\ldots$.}
\label{fig-cage}
\end{figure}

The term $\mu^{\rm eff}_{\rm born}$ represents the Born term \eqref{eq-mu-born} associated with
the renormalized potential \eqref{eq-veff}, which itself consists of the
original born term \eqref{eq-mu-born} at $O(A^{0})$ and corrections at $O(A)$.
The last two terms on the r.h.s of
\eqref{eq-hat-mu-harmonic} represents the effect of 
the stress relaxation due to the fluctuation inside
the cages, i.e. $\beta$-relaxation (See Fig.~\ref{fig-cage}).

\begin{figure}[htbp]
%\begin{center}
\includegraphics[width=0.45\textwidth]{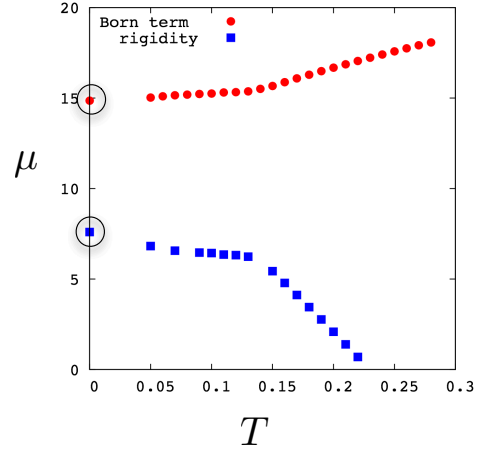}
%\end{center}
\caption{The rigidity of a binary soft-core system. 
The born term which represents the instantaneous, affine response
and the rigidity $\hat{\mu}$ which includes non-affine corrections by
stress relaxation due to fluctuations inside cages are shown.
The results are $T>0$ are reported in \cite{paper1,paper2}. 
The Kauzmann temperature is $T_{\rm K}/\epsilon \sim 0.12$.
}
\label{fig-shearmodulus-softcore}
\end{figure}

We show the result of a model computation on the binary soft-core system
$v(r) \sim (r/a)^{-12}$ in Fig.~\ref{fig-shearmodulus-softcore}. 
The result at finite temperature $T>0$ was already reported in 
\cite{paper1,paper2}. Here we added the result of a computation
performed directly in the zero temperature limit $T \to 0$.

The cloned liquid computation can be performed
in the zero temperature limit $T \to 0$ by introducing scaled variables
$\alpha=A/T$ and $\tau=T/m$ \cite{Berthier-Jacquin-Zamponi-2011}.
The expression of the rigidity $\hat{\mu}$ given \eq{eq-hat-mu-harmonic} 
also suggest it has a limiting value in the $T \to 0$ limit.

For the binary soft-core system we consider $T \to 0$ limit of the
the theory formulated in Ref \cite{coluzzi-mezard-parisi-verrocchio-1999}
which employs the binary HNC approximation for the computation of the
liquid free energy and the radial distribution function $g_{T}(r)$.
As the result we find the effective temperature of the
cloned liquid converges to $\tau \equiv \lim_{T \to 0} T/m^{*}(T) \simeq
0.115$ within the 1st order cage expansion. This value is close to the
Kauzmann transition temperature $T_{\rm K}/\epsilon \sim 0.12$ where $m^{*}(T_{\rm
K})=1$ \cite{coluzzi-mezard-parisi-verrocchio-1999}. 

 The result shown in
Fig.~\ref{fig-shearmodulus-softcore} suggest 
the rigidity at low temperatures $T < T_{\rm K}$
behaves essentially as if at $T=0$. 
Moreover the numerical values of the plateau modulus $G_{\rm p}$ observed by MD simulations 
\cite{barrat-roux-hansen-klein-1988,okamura-thesis},
and the value of the rigidity of the inherent structures
\cite{yoshino-lemaitre-2012} compare well with the theoretical values shown in
Fig.~\ref{fig-shearmodulus-softcore}.
These observations strongly support the usual view
that the metastable glassy states at low enough temperatures
are harmonic solids which can be efficiently described as systems of random spring network (see also \cite{harrowell-2012} for a related work). 

Note also that the rigidity becomes significantly smaller 
with increasing temperature above $T_{\rm K}$.
We interpret this as reduction of the rigidity \cite{paper2}
due to thermally activated plastic events 
among a union of inherent structures
\cite{Stillinger-Weber-1982} belongin to a common metabasin
\cite{Doliwa-Heuer-2003} or a metastable state. 
The rigidity apparently vanishes (crosses $0$) around $T/\epsilon \sim 0.22$, suggesting
melting of metastable states, which happens to be rather close to 
the so called MCT critical temperature $T_{\rm c}/\epsilon \sim 0.19-0.22$
\cite{roux-barrat-hansen-1989}\cite{berthier-private-communication-2012}.
However the first order cage expansion which we have employed does not allow us
to locate the MCT transition temperature at which the glassy solution
with finite cage size $A < \infty$ disappear presumably 
by a spinodal like mechanism \cite{Franz-Parisi}. 
As we noted in \cite{paper2} we rather consider at the 
moment that, at least in the mean-field sense, 
the rigidity should exhibit a discontinuous behaviour.
This is because the expression \eqref{eq-hat-mu-harmonic} actually implies
that the rigidity is a function of the the cage size $A$
which is predicted to behave as
$A(T)-A(T_{\rm c}) \propto \sqrt{T_{\rm c}-T}$ approaching the dynamical
temperature $T_{\rm c}$ 
from below by the mode coupling theory (MCT) \cite{MCT}.
Indeed such a discontinuous behaviour of the rigidity has been suggested
in an alternative formulation of the replica approach\cite{szamel-flenner}.

\section{Repulsive contact systems}

Finally we are in the position to analyze the rigidity of the glassy states of 
repulsive contact systems: the hard-spheres and generic soft repulsive
contact systems in the low temperature limit $T \to 0$. 

Again we first consider the rigidity matrix of the 'free' $m$-replica 
system, which is obtained as,
\begin{eqnarray}
&& \beta \mu_{ab} = V \left [  \sum_{\langle kl\rangle} \langle \beta\sigma({\bf r}^{a}_{kl})\rangle
\langle \beta \sigma_{b}({\bf r}^{b}_{kl})\rangle  \right.  \nonumber \\
&&  \left. -\sum_{\langle kl \rangle}\sum_{\langle mn \rangle \neq \langle kl \rangle} 
(\langle \beta \sigma({\bf r}^{a}_{kl}) \beta \sigma({\bf r}^{b}_{mn})\rangle 
%\right. \nonumber \\
%&& \left. 
-\langle \beta \sigma(r^{\bf a}_{kl})\rangle\langle \beta\sigma(r^{b}_{mn})\rangle)
\right]\nonumber.
\end{eqnarray}
As expected it is similar to the expression \eq{eq-mu-hardsphere} for the single system.

Then by switching on the cloning we obtain 
the rigidity of metastable states (plateau modulus $G_{\rm p}$) up to $O(A)$ as,
\begin{eqnarray}
\beta \hat{\mu} =  -\frac{1}{m^{*}}
\left (\frac{A^{*}}{m^{*}} \right) 
\frac{6}{\pi}\frac{\phi}{\sigma^{3}} \int d^{3}r_{1}d^{3}r_{2} 
\nonumber \\
\left ( \nabla \beta m^{*} \sigma({\bf r}_{1}) \right)
\cdot  \left (\nabla \beta m^{*} \sigma({\bf r}_{2}) \right) 
(g_{3})_{T/m*,\phi}({\bf r}_{1},{\bf r}_{2}) 
\label{eq-hat-mu-hardsphere}
\end{eqnarray}
Most important difference from the case of the harmonic systems
\eqref{eq-hat-mu-harmonic} is that the Born term is apparently
absent here. Consequently the rigidity is $0$ at order $O(A^{0})$ 
and starts only at $O(A)$. The relevant term at $O(A)$ is
again the one related to the thermal fluctuation of the shear-stress
due the the fluctuations inside cages (See Fig.~\ref{fig-cage}).

In order to make further progresses, 
we approximate the three-point correlation function $g_{3}({\bf r}_{1},{\bf r}_{2})$ 
by the Kirkwood approximation,
\beq
(g_{3})_{T,\phi}({\bf r}_{1},{\bf r}_{2}) \simeq g_{T,\phi}(r_{1})
g_{T,\phi}(r_{2})g_{T,\phi}(|{\bf r}_{1}-{\bf r}_{2}|).
\eeq
Then using the cavity function $y_{T,\phi}(r)$ defined as 
$g_{T,\phi}(r)\equiv y_{T,\phi}(r)e^{-\beta v(r)}$ we find,
\beq
\beta \hat{\mu} =  \frac{1}{m^{*}}
\left (\frac{A^{*}}{m^{*}} \right)
\frac{6}{\pi}\frac{\phi}{\sigma^{3}} C
\label{eq-hat-mu-hardsphere-2}
\eeq
with
\begin{eqnarray}
&& C \simeq -\int d^{3}r_{1}d^{3}r_{2} 
y_{T/m^{*},\phi}(|{\bf r}_{1}-{\bf r}_{2}|)e^{-\beta m^{*}v(|{\bf r}_{1}-{\bf r}_{2}|)} \nonumber \\
&&
\nabla_{1}\left(y_{T/m^{*},\phi}(r_{1})\hat{z}_{1}\hat{x}_{1} r_{1}\frac{d e^{-\beta m^{*}v(r_{1})}}{dr_{1}}  \right) \nonumber \\
&& \cdot \nabla_{2}\left(y_{T/m^{*},\phi}(r_{2})\hat{z}_{2}\hat{x}_{2} r_{2}\frac{d e^{-\beta m^{*}v(r_{2})}}{dr_{2}} \right) \nonumber \\
&& \xrightarrow[T \to 0]{}
-(y^{\rm HS}_{\phi}(a))^{2}
\int d\Omega_{1}d\Omega_{2}\int dr_{1}d r_{2} r_{1}^{2}r_{2}^{2} \nonumber \\
% \int d^{3}r_{1}d^{3}r_{2} 
&& y^{\rm HS}_{\phi}(|{\bf r}_{1}-{\bf r}_{2}|) \theta(|{\bf r}_{1}-{\bf r}_{2}|-a)\nonumber \\
&& 
\nabla_{1}\left(\hat{z}_{1}\hat{x}_{1} r_{1}\delta(r_{1}-a)  \right) 
\cdot \nabla_{2}\left(\hat{z}_{2}\hat{x}_{2} r_{2}\delta(r_{2}-a)  \right) 
\label{eq-C}
\end{eqnarray}
In the last equation we took $T \to 0$ limit which greatly simplifies 
the calculation.
The evaluations of the integrals over the polar coordinates
are tedious but straight forward. First the integrations along 
the radial coordinates $r_{1}$ and $r_{2}$ can be done exactly
via integrations by parts. Subsequently 
the integrations over the solid angles $\Omega_{1}$
and  $\Omega_{2}$ can be simplified, with the help of spherical harmonics,
to simple one dimensional integrals over $x=\cos(\theta_{12})$ where
$\theta_{12}$ is the relative angle between the two solid angles.
If we make a further approximation 
$y^{\rm HS}_{\phi}(|{\bf r}_{1}-{\bf r}_{2}|) \simeq y^{\rm HS}_{\phi}(a)$
the last integrations can also be done exactly and we finally obtain,
\beq
C \simeq (y^{\rm HS}_{\phi}(a))^{3} \frac{113}{120}\pi^{2}.
\label{eq-C-2}
\eeq
The above results suggest finite rigidity $\hat{\mu}$ 
of the repulsive contact systems which is proportional to the temperature $T$
meaning that it is of entropic origin in sharp contrast to the harmonic
systems discussed previously whose rigidity is essentially mechanical.

In order to study how the rigidity depends
on the volume fraction $\phi$, we need to know the values
of $m^{*}(T=0,\phi)$, $A^{*}(T=0,\phi)$ and $y^{\rm HS}_{\phi}(a)$.
Fortunately they are provided by recent studies on the hard-sphere glass \cite{parisi-zamponi-2010} and a soft repulsive contact potential system \cite{Berthier-Jacquin-Zamponi-2011}
close to the jamming density, more precisely the so called glass close packing density
$\phi_{\rm GCP}$. According to the latter works 
$m^{*}(T=0,\phi)\simeq c_{1}(\phi_{\rm GCP}-\phi)$
and $A^{*}(T=0,\phi)\simeq c_{2}(\phi_{\rm GCP}-\phi)$
approaching $\phi_{\rm GCP}$ from below with $c_{1}$ and $c_{2}$ being some positive constants.
The value of $y^{\rm HS}_{\phi_{\rm GCP}}(a)$ is also positive.
Using these information in our result we obtain,
\beq
\lim_{T \to 0} \beta \hat{\mu} =\frac{c}{\phi_{\rm GCP}-\phi} 
\label{eq-scaling-hat-mu}
\eeq
with the numerical pre-factor given by,
\beq
c= 6 \pi
\frac{\phi_{\rm GCP}}{\sigma^{3}}
\frac{c_{2}}{c_{1}^{2}}\frac{113}{120}
 (y^{\rm HS}_{\phi_{\rm GCP}}(a))^{3}.
\label{eq-prefactor-c}
\eeq
Using the numerical values of constants reported in \cite{Berthier-Jacquin-Zamponi-2011}
we find $c \sim 0.7$.
A remarkable feature is that the scaling of the rigidity $\hat{\mu}$ 
found above \eq{eq-scaling-hat-mu} is exactly the same as that of
the pressure $p$ found in \cite{parisi-zamponi-2010} 
and \cite{Berthier-Jacquin-Zamponi-2011}. 
We note that our result is different from that of
Ref \cite{Brito-Wyart-2006} which predicts
somewhat stronger rigidity $\hat{\mu} \propto p^{3/2}$ 
based on an effective (yet microscopic) 
harmonic description developed for the inherent structures of the hardsphere glass.

The scaling \eqref{eq-scaling-hat-mu} agrees with
the plateau modulus $G_{\rm p}$ observed by MD simulations
of the stress relaxation on the same system \cite{okamura-thesis,okamura-yoshino-2012}.
However the evaluation of the numerical factor $c$ \eqref{eq-prefactor-c}
may be improved in several respects: 1) the approximation
$y^{\rm HS}_{\phi}(|{\bf r}_{1}-{\bf r}_{2}|) \simeq y^{\rm HS}_{\phi}(a)$
can be avoided by doing numerical integrations 2) higher order corrections
terms of the cage expansion due to renormalization of the potential 
\cite{parisi-zamponi-2010,Berthier-Jacquin-Zamponi-2011} can be considered
3) better evaluation of the 3-point correlation function $g_{3}({\bf
r}_{1},{\bf r}_{2})$ than the Kirkwood approximation may be considered. 
We have checked that the item 2) 
amounts to reduction of the value of $c$ by an amount of about $20$\% 
which will be reported elsewhere. The item 3) would be challenging but
worthwhile.

It is interesting to compare the above result with
some experimental observations of the rigidity of
densely packed repulsive colloids \cite{Mason-Weitz-1995} 
and emulsions \cite{Mason-Bibette-Weitz-1995,Weitz97}.
The experiments were performed at the room temperature
which is actually a very low temperature for these systems.
For example the reduced temperature can be estimated as
$k_{\rm B}T/\epsilon \sim 10^{-5}$ for the
emulsions system \cite{Mason-Bibette-Weitz-1995,Weitz97}.
However the experimental data reveal presence of 
finite entropic rigidity which rapidly increase approaching the
jamming density from below. A striking feature found
by the experiment on the emulsion system is that simultaneous measurement 
of the pressure reveals that the pressure 
and the shear-modulus behave very similarly  (See Fig. 3 of
\cite{Mason-Bibette-Weitz-1995}).  Thus it appears that the our
theoretical result is consistent with the experiment.

In the present paper we have limited ourselves to the volume fractions
$\phi < \phi_{\rm GCP}$, but it is straightforward to extended the
present approach to the jammed region $\phi > \phi_{\rm GCP}$ 
concerning the systems of the soft repulsive contact potentials. 
It is important to note that for this class of systems one cannot 
rely on the usual picture of harmonic solids naively \cite{Schreck-et-al-2011,Wyart-2012,ikeda-berthier-biroli-2012}.
We will report the results elsewhere with detailed comparisons with the 
known results \cite{Durian-1995,Mason-Bibette-Weitz-1995,Weitz97,Nagel-group,Wyart-2005,Brito-Wyart-2006,hecke-2010}.

\section{Conclusions}

In the present paper we first reviewed a microscopic approach 
to study the rigidity of structural glasses
or the plateau modulus $G_{\rm p}$ of supercooled liquids based on the 
cloned liquid approach. 
Then we discussed how to extend the method, which has been
limited to the cases of harmonic glasses, i.~e. systems with analytic
potentials, to the cases of the glassy repulsive contact systems
like the hardsphere glasses or soft repulsive contact systems in the low
temperature limit. We found the entropic rigidity of this class of 
systems exhibit divergent behaviour much as the pressure 
approaching the jamming density
from below, which appear to be consistent with experimental observations
on repulsive colloids and emulsions.

%%%%%%%%%%%%%%%%%%%%%%%%%%%%%%%%%%%%%%%%%%%%%%%%
%% BACKMATTER
%%%%%%%%%%%%%%%%%%%%%%%%%%%%%%%%%%%%%%%%%%%%%%%%

\begin{theacknowledgments}
The author thanks Marc M\'{e}zard, Satoshi Okamura, 
Ana\"el Lema\^{\i}tre,
Hugo Jacquin and Francesco Zamponi for useful discussions.
This work is supported by a {\it Triangle de la physique} grant number 117 "Intermittent response of glassy systems at mesoscopic scales",
and  Grant-in-Aid for Scientific Research (C) (50335337).
\end{theacknowledgments}



\bibliographystyle{aipproc}   % if natbib is available
%%\bibliographystyle{aipprocl} % if natbib is missing
%
%%%%%%%%%%%%%%%%%%%%%%%%%%%%%%%%%%%%%%%%%%%%
%%% You probably want to use your own bibtex database here
%%%%%%%%%%%%%%%%%%%%%%%%%%%%%%%%%%%%%%%%%%%%
%\bibliography{sample}
%
%%%%%%%%%%%%%%%%%%%%%%%%%%%%%%%%%%%%%%%%%%%%
%%% Just a reminder that you may have to run bibtex
%%% All of it up to \end{document} can be removed
%%% if you don't like the warning.
%  \typeout{******************************************}
%  \typeout{** Please run "bibtex \jobname" to optain}
%  \typeout{** the bibliography and then re-run LaTeX}
%  \typeout{** twice to fix the references!}
%  \typeout{******************************************}

%%%%%%%%%%%%%%%%%%%%%%%%%%%%%%%%%%%%%%%%%%%
%% The following lines show an example how to produce a bibliography
%% without the help of the BibTeX program. This could be used instead
%% of the above.
%%%%%%%%%%%%%%%%%%%%%%%%%%%%%%%%%%%%%%%%%%%

\end{document}